\pdfoutput=1
\documentclass[journal]{IEEEtran}
\ifCLASSINFOpdf
   \usepackage[pdftex]{graphicx}
\else
\fi
\usepackage{graphicx}
\usepackage{epstopdf}

\usepackage{bibspacing}
\usepackage{algpseudocode}
\usepackage{algorithm}
\usepackage{array}
\usepackage{multirow}

\usepackage{xcolor}

\usepackage{textcomp}
\usepackage[noadjust]{cite}
\usepackage{wrapfig}

\begin{document}

%
\title{A Novel Traffic Rate Measurement Algorithm for QoE-Aware Video Admission Control}
%
%
%

\author{Qahhar~Muhammad~Qadir, Alexander~A.~Kist, and Zhongwei~Zhang
\thanks{Manuscript received August 09, 2014; revised December 19, 2014 and March 10, 2015; accepted March 21, 2015. Date of publication March 25, 2015.

Qahhar Muhammad Qadir and Alexander A. Kist are with the School of Mechanical and Electrical Engineering, University of Southern Queensland, Toowoomba, QLD 4350, Australia (e-mail: safeen.qadir@ieee.org; kist@ieee.org).}
\thanks{Zhongwei Zhang is with the School of Agricultural, Computational and Environmental Sciences, University of Southern Queensland, Toowoomba, QLD 4350, Australia (e-mail: zhongwei.zhang@usq.edu.au).
	
\noindent \textcolor{blue}{$\copyright$ 2015 IEEE. Personal use of this material is permitted. Permission from IEEE must be obtained for all other uses, in any current or future media, including reprinting/republishing this material for advertising or promotional purposes, creating new collective works, for resale or redistribution to servers or lists, or reuse of any copyrighted component of this work in other works.} \textcolor{red}{Digital Object Identifier 10.1109/TMM.2015.2416637}
}
}
\maketitle

\begin{abstract}
With the inevitable dominance of video traffic on the Internet, providing perceptually good video quality is becoming a challenging task. This is partly due to the bursty nature of video traffic, changing network conditions and limitations of network transport protocols. This growth of video traffic has made Quality of Experience (QoE) of the end user the focus of the research community. In contrast, Internet service providers are concerned about maximizing revenue by accepting as many sessions as possible, as long as customers remain satisfied. However, there is still no entirely satisfactory admission algorithm for flows with variable rate. The trade-off between the number of sessions and perceived QoE can be optimized by exploiting the bursty nature of video traffic. This paper proposes a novel algorithm to determine the upper limit of the aggregate video rate that can exceed the available bandwidth without degrading the QoE of accepted video sessions. A parameter $\beta$ that defines the exceedable limit is defined. The proposed algorithm results in accepting more sessions without compromising the QoE of on-going video sessions. Thus it contributes to the optimization of the QoE-Session trade-off in support of the expected growth of video traffic on the Internet.

\end{abstract}


%
\IEEEpeerreviewmaketitle

\section{Introduction}
%
%
%
%

\IEEEPARstart{W}{ith} the inevitable dominance of video traffic on the Internet, it is becoming a challenging task to provide perceptually good video quality. This is partly due to the bursty nature of video traffic, changing network conditions and limitations of network transport protocols. Cisco predicts that ``The sum of all forms of video (TV, video on demand [VoD], Internet, and P2P) will be in the range of 80 to 90 percent of global consumer traffic by 2018'' \cite{Cisco2014a}. Over the last decade, efforts have been made to provide Quality of Service (QoS) within the core network by considering objective parameters at the network layer such as bandwidth, delay and jitter. Diffserv \cite{BlakeS.1998} is an example of these paradigms that can support QoS. The research community and Internet Service Providers (ISP)s have made subjective quality, as perceived by the end user, a main research target. The International Telecommunication Union (ITU) defines this parameter as ``Quality of Experience'' (QoE) \cite{ITU-T2007}. The current design of the Internet has to be enhanced to extend the scope of QoS to consider end-to-end quality, be content-aware and user centric.

Admission control is a well known technique to keep traffic load at acceptable levels and guarantee quality for admitted sessions via resource reservation. This idea has been adopted in the past in QoS architectures such as in Diffserv. Thus, some sort of explicit admission control is required to provide per-session QoE by which the network has the right to deny sessions to ensure that the QoE of active sessions is not affected by new sessions. ISP are concerned about maximizing revenue by accepting as many sessions as possible. Measurement-Based Admission Control (MBAC) has been proposed as a solution. In contrast to parameter-based admission control, it is better suited to video traffic. MBAC relies on the measurement of video characteristics such as current load and peak rate. Different algorithms have been proposed to estimate network load \cite{Breslau2000}; however there are algorithms which rely on the Instantaneous Aggregate Arrival Rate (\emph{IAAR}) for their operations.

Despite all the efforts, there is no entirely satisfactory admission algorithm for variable rate flows \cite{Auge2011}. Admission control algorithms must not rely on worst-case bounds or instantaneous video arrival rate, as they do not reflect the bursty characteristic of video traffic. This is due to the fact that the burstiness of video flows can be compensated by the silence of other flows. The Internet Engineering Task Force (IETF) has standardized the Pre-Congestion Notification (PCN) based admission control for the Internet \cite{Menth2012} which merely relies on the calculated rate for a measurement period. The perceived QoE-Session relationship can be greatly optimized by exploiting the bursty nature of video traffic.

This paper contributes to the measurement mechanism for QoE-aware admission control. It proposes a novel traffic rate measuring algorithm for video admission control mechanisms. The relationship between \emph{IAAR} and the proposed rate is established mathematically. We call the proposed measured rate ``Proposed Instantaneous Aggregate Arrival Rate'' (\emph{Pro-IAAR}) and proposed admission control procedure based on \emph{Pro-IAAR} ``\emph{Pro-IAAR}-Based Measurement Admission Control'' (\emph{Pro-IBMAC}). We also call the admission control procedures which are based on the Calculated Rate (\emph{CalR}) such as PCN ``\emph{CalR}-Based Admission Control'' (\emph{CBAC}).

Whereas traffic measurement algorithms and MBAC have been widely covered by the research community, to the best of our knowledge this is the first work that includes QoE in the area of the QoE-Session optimization. The main contributions of this paper are twofold:

\begin{enumerate}
\item A novel algorithm for traffic measurement supported by the mathematical model is proposed. The algorithm measures the exceedable video aggregate rate that is able to keep the video quality unimpaired. The exceedable rate is the total bitrate of enrolled video traffic that can exceed the available link capacity without degradation to the user's perception of quality.

\item Operation of the proposed measurement algorithm is demonstrated with an implementation in a QoE-aware admission control procedure for video admission.
\end{enumerate}

The remainder of the paper is organized as follows. Section \ref{sec:relatedWork} presents related work. Assumptions made by this paper are detailed in Section \ref{sec:assumption}. Section \ref{sec:QoE} provides a theoretical background of QoE. Section \ref{sec:model} presents the mathematical model for the proposed algorithm. The simulation setup is explained and results are presented and discussed in Section \ref{sec:resultANDdiscuttion}. Section \ref{sec:subjectiveTest} describes the environment of the subjective tests and analysis of the collected data. The proposed model is validated in Section \ref{sec:validation}. The paper concludes with Section \ref{sec:conclusion}.

\section{Related Work}
\label{sec:relatedWork}
MBAC is not a new topic as work has been undertaken since video traffic has emerged on the Internet. It includes two main components: measurements of network load and admission policies. Four MBAC algorithms are presented in \cite{Gibbens1997} based on Chernoff bounds. A MBAC scheme based on measured mean and variance of load offered to the cross-protect priority queue is proposed in \cite{Auge2011}.

As traffic flow rate is only meaningful if it is associated with a corresponding interval length. Network traffic over some interval has been studied as an essential part of the MBAC functionality. The admission control scheme proposed in \cite{Floyd1996} estimates the equivalent capacity of a class of aggregated traffic based on Hoeffding bounds for controlled-load services. The suitability of the average instead of the instantaneous arrival rate for video streaming admission decisions has been investigated in \cite{Qadir2013a}. An algorithm for MBAC has been introduced in \cite{Qiu2001} that employs adaptive and measured peak rate envelopes of the aggregate traffic flow to allocate resources for multiclass networks with link sharing. The flow behavior as a function of interval length can be described by the proposed rate envelope which characterizes the extreme values (maximal rates) of the aggregate flow that can avoid packet loss. As a supporting mechanism in flow and admission control, techniques have been developed for estimating available bandwidth \cite{Nam2012}, \cite{Guerrero2010}, \cite{Nam2013}, \cite{Lubben2014} and \cite{Cavusoglu2014}.

Other studies have compared the performance of MBAC algorithms. The simple sum; a parameter-based admission control algorithm has been compared to three measurement-based algorithms; the measured sum, acceptance region and equivalent bandwidth based on the link utilization and adherence to service commitment \cite{Jamin1997}. The robustness of \cite{Floyd1996}, \cite{Jamin1997} and \cite{Qiu2001} in meeting the QoS target have been compared in \cite{Nevin2008}. They have been further evaluated based on maximum tolerable packet loss rate and maximum packet queuing delay without assuming any explicit knowledge on incoming flows and on-going traffic \cite{Ammar2011}. All of the three studied algorithms were found to meet the first target of maximum tolerable packet loss rate while only \cite{Qiu2001} was able to always meet the second target of maximum packet queuing delay. The knowledge-base admission control scheme introduced in \cite{Ammar2012} determines whether to accept a flow based on QoS performance parameters such as maximum tolerable delay or packet loss rate. The scheme achieves a good trade-off between flow performance and resource utilization compared to \cite{Jamin1997} and \cite{Qiu2001}. In \cite{Lima2007} the architecture of centralized, distributed, hybrid, class-based and active/passive MBAC and their limitations on the quality control of network services have been compared.

The efficiency of MBAC algorithms depends on interactions between several time-scales, ranging from the very short time scales to the entire session. Work in \cite{Nevin2010} has studied how uncertainty in the measurements of MBAC varies with the length of the observation window and described a methodology for analyzing measurement errors and performance. The concept of similar flows and adding slack in bandwidth have been introduced to minimize the probability of false acceptance. In \cite{Moore2002} an implementation-based comparison of MBAC algorithms has been made using a purpose built test environment. It has revealed that there is no single ideal MBAC algorithm due to computation overheads, multiple timescales present in both traffic and management and error resulting from random properties of measurements. These features dramatically impact the MBAC algorithm's performance.

Work presented in \cite{Nam2008} has proposed a delay-aware admission control to guarantee delay bounds for delay sensitive applications. The video quality model presented in \cite{Zhang2013} targets Skype video calls based on measurement and can be used for user QoE-aware network provisioning. The model can find the minimum bandwidth needed to accommodate \emph{N} concurrent Skype video calls with satisfactory Mean Opinion Score (MOS). The study conducted in \cite{Xu2014} has investigated the system architecture, video generation and adaptation, packet loss recovery and QoE of video-conferencing solutions. Google+, iChat and Skype were all covered in the study. The delivered quality was measured in terms of end-to-end delay in a wide range of real and emulated network scenarios. The study has found that the layered video coding and server architecture (used by Google+ and Skype) can significantly improve user conferencing experiences. Most recently, \cite{ChendebTaher2014} has proposed a model-based admission control algorithm to predict the QoS metrics based on which and the QoS constraints of the flows, appropriate decision for new flow is taken. The average number of satisfied users was maximized through a QoE-aware scheduling framework by sending a single bit feedback to indicate the satisfaction level \cite{Lee2014}.

As a cutting edge proposed admission control mechanism for multimedia network, the PCN-based admission control \cite{EardleyP.2009} has attracted the attention of researchers. Several modifications to the PCN algorithm have been proposed in \cite{Latre2011b}. An extension to the PCN-based admission control system has been proposed in \cite{Latre2011a}. A novel metering algorithm based on a sliding-window, to cope with the bursty nature of video sessions and another adaptive algorithm to facilitate the configuration of PCN were proposed in that work.

Admission control has also been proposed to better support applications with QoS requirements in wireless networks. The appropriate thresholds for admission decisions were studied by \cite{Xu2013}. A flow-level mechanism for multiple antenna equipped nodes to maximize flow acceptance and improve network throughput has been developed in \cite{Hamdaoui2007}. A QoE-based admission control for wireless has been proposed in \cite{Piamrat2008} in which the access point controls video sessions based on the MOS scores computed by pseudo-subjective quality assessment tool run on the access point.

Most of the MBAC algorithms that have been discussed in the literature are per-aggregate MBAC algorithms. The per-flow MBAC algorithm presented in \cite{Jiang2005} targets the flow-aware network by adopting dynamic priority scheduling for flow aggregation. A newly admitted flow is given a lower priority by the proposed algorithm, however its priority is improved when an existing flow leaves the network. Finally, an enhancement to the MBAC has been proposed to mitigate the impact of fair rate degradation and ensure better quality in flow-aware network by \cite{Wojcik2013}.

\section{Assumptions}
\label{sec:assumption}

\noindent This paper makes the following assumptions:
\begin{itemize}

\item Video traffic is the dominant Internet traffic \cite{Cisco2014a}. It is the only traffic that is subject to admission control. Other traffic volumes are small in comparison and therefore only video traffic is considered.

\item Video traffic is bursty in nature as video applications in reality send traffic at a very variable rate \cite{Nevin2010}.

\item An explicit admission control is required to provide an acceptable level of QoE \cite{Nevin2010} on bottleneck links.

\item There is no danger of a ``flash crowd'', in which many admission requests arrive within the reaction time of admission mechanisms, because then they all might get admitted and so overload the network \cite{EardleyP.2009}.

\end{itemize}

\section{QoE; A New Quality Paradigm}
\label{sec:QoE}

QoE is the quality as experienced by end users. The purpose of introducing QoE is to include all aspects of multimedia systems that are related to media quality. Addressing quality from end user experience or perceived QoE is a relatively new approach which requires more research in all directions such as optimization, assessment, monitoring, management and prediction. This is due to the emergence of massive video services and development of huge number of video capable devices such as smart phones \cite{Cisco2014a}. Various layers (from video encoding to decoding) and across the access and/or core networks are involved in providing an end-to-end QoE to end users. Technically, perceived video quality is affected by the trade-off relationship between encoding redundancy and network impairments. In addition to network parameters such as bandwidth, delay, packet loss ration, other technical and non-technical parameters may affect quality \cite{Brooks2010}.

There are different approaches to measure and estimate QoE. Subjective, objective or hybrid approaches are mainly used for that purpose. Since people have different perceptions of the same video content, groups of people carry out subjective tests by grading the shown sequence. This is time-consuming and costly; however it is worthwhile as real users are involved in the tests. Objective video quality metrics are often proposed because none of the QoS parameters can precisely define the QoE of multimedia services \cite{Latre2009}. These objective approaches are carried out by the use of algorithms and formulas. Peak Signal to Noise Ratio (PSNR) and Structural Similarity (SSIM) are two full reference objective video quality metrics. They compare the original video with received (possibly distorted) video and calculate the MOS value. PSNR is mostly used for its simplicity and good correlation with the subjective video test result. PSNR tools are available to calculate the PSNR value. A possible mapping of PSNR to MOS is shown in Table \ref{tab:psnr_mos_mapping} \cite{Ohm2004}. However, this is a problematic approach as PSNR does not directly correspond to MOS \cite{Gross2004}. On the other hand, SSIM estimates the perceived quality frame by frame and is considered to have a higher correlation with subjective quality ratings \cite{Group2008}. The SSIM index assumes that the human visual system is more oriented towards the identification of structural information in video sequences. It produces a score between 0 and 1 from original and received signals \cite{Wang2004a}. The third approach is a hybrid between subjective and objective methods in which both the technical parameters as well as human rating are taken into account \cite{Cherif2011} \cite{Piamrat2009a}. ITU recommends objective modeling of measurable technical performance and subjective testing with people \cite{Brooks2010}.

\begin{table}[!th]
			\centering
			\caption{Possible PSNR to MOS mapping}
			\label{tab:psnr_mos_mapping}
        	\begin{tabular}{|l|c|c|}
			\hline
			PSNR	&	MOS	&	Quality	\\
			\hline
			$>$37 		&	5	&	Excellent \\
			\hline
			31-37 	&	4	&	Good \\
			\hline
			25-31 	&	3	&	Fair \\
			\hline
			20-25 	&	2	&	Poor \\	
			\hline
			$<$20 	&	1	&	Bad \\
			\hline
			\end{tabular}
\end{table}

\section{Modeling}
\label{sec:model}
\subsection{Measurement Algorithm}
\label{sec:measurmentAlgorithm}

In this section, we describe a new approach to measure traffic rate that suits video traffic. For the benefit of comparison, we introduce the traditional approach of traffic measurement \emph{IAAR} then present our proposed measurement algorithm \emph{Pro-IAAR}. Since, the measurement mechanism is proposed for video admission procedures, it will be modeled into an admission control scheme \emph{Pro-IBMAC}; Equation (\ref{eq:9}). \emph{IAAR} at any time \emph{t}$>$0 and \emph{i}$>$0 can be expressed by Equation (\ref{eq:1}):

\begin{equation}
\label{eq:1}
IAAR(t) = \sum_{i=1}^n x_{i}(t)
\end{equation}

\noindent where \emph{$x_i(t)$} is the instantaneous arrival rate (throughput) of session \emph{i} at time \emph{t}, and \emph{n} is the number of sessions. Let 
\emph{$x_i(t)$} be an independent random variable with minimum rate \emph{$x_i^{min}(t)$}, peak rate \emph{$x_i^{max}(t)$} and \emph{$x_i^{min}(t)$} $\le$ \emph{$x_i(t)$} $\le$ \emph{$x_i^{max}(t)$}. Further assuming that \emph{$x_i(t)$} is a discrete random variable that takes any set of values from a finite data set \emph{$x_1(t)$}, \emph{$x_2(t)$}, .... \emph{$x_n(t)$} each of probability \emph{$p_1(t)$}, \emph{$p_2(t)$}, .... \emph{$p_n(t)$} respectively.

A new session will be accepted by \emph{CBAC}, if the sum of \emph{CalR($\tau$)} for the time window ($\tau$) plus the peak rate of the new session \emph{$x_{new}$} is less or equal to the link's capacity \emph{$C_{l}$} as given by Equation (\ref{eq:2}):

\begin{equation}
\label{eq:2}\
CalR(\tau) + x_{new} \le  C_{l}
\end{equation}

In our proposed scheme we consider \emph{Pro-IAAR(t)} as an admission parameter instead of \emph{CalR($\tau$)}. Now we find how \emph{Pro-IAAR(t)} is related to \emph{IAAR(t)}. We utilize the Hoeffding inequality theorem \cite{Hoeffding1963} to develop a model for \emph{Pro-IAAR(t)}. The reason behind this approach is that the Hoeffding theorem relates \emph{IAAR(t)} and the average of \emph{IAAR(t)}; $\mu_S(t)$. It defines the upper bound of the probability that the sum of \emph{n} independent random variables will be greater than the average by \emph{n}$\epsilon$ or more for $\epsilon$ $>$ \emph{0}. Equation (\ref{eq:3}) quantifies this probability relationship between \emph{IAAR(t)} and $\mu_S(t)$. We then develop a relationship between \emph{Pro-IAAR(t)} and \emph{IAAR(t)}. Hoeffding bounds were first used for admission control algorithms in \cite{Floyd1996}.

\begin{equation}
\label{eq:3}
Pr\{IAAR(t) \ge \mu_S(t)+n\epsilon\} \le \gamma
\end{equation}

\noindent where $\gamma$ is given by Equation (\ref{eq:4}):

\begin{equation}   
\label{eq:4}\
\gamma = exp \left(   \frac{-2n^2\epsilon^2} { \sum_{i=1}^n (x_{i}^{max}(t)-x_{i}^{min}(t))^2} \right)  
\end{equation}

\noindent $\mu_S(t)$ is the expectation value of \emph{IAAR(t)} which is given by Equation (\ref{eq:5}) in which \emph{$p_{i}$} represents the probability the session \emph{i} is active:

\begin{equation}   
\label{eq:5}\
\mu_S(t) = EIAAR(t) = \sum_{i=1}^n x_{i}(t)\  p_{i}(t)
\end{equation}

The term $\mu_S(t)+n\epsilon$ in Equation (\ref{eq:3}) represents the proposed \emph{Pro-IAAR(t)} at time \emph{t} which is given by Equation (\ref{eq:6}) and $\epsilon$ is given by Equation (\ref{eq:7}):

\begin{equation}
\label{eq:6}
Pro\textit{-}IAAR(t) = \mu_S(t)+n\epsilon
\end{equation}

\begin{equation}   
\label{eq:7}\
\epsilon = \beta \mu_S(t) \frac{n-1}{n} \ \ \ \ \ \ \ \ \ \ \ \ \ \ 0<\beta\le1
\end{equation}

Parameter $\beta$ represents how much the total bitrate of enrolled video traffic can exceed the available link capacity without degradation to the user perception quality. It governs the degree of the efficiency of \emph{Pro-IBMAC}. Therefore, choosing a proper value for $\beta$ controls the degree of risk of the admission decision as it balances the QoE-Session trade-off relationship. The value of $\beta$ that optimizes this relationship is referred to as ``proposed value'' in this paper.

The condition $\epsilon$$>$0 of Equation (\ref{eq:3}) is satisfied by setting $\beta>$0 in Equation (\ref{eq:7}) (assuming that \emph{n}$>$1). Although Equation (\ref{eq:7}) is also valid for $\beta>$1, the scope of the proposed scheme is only for 0$<\beta\le$1. High values of $\beta$ within this range lets \emph{Pro-IBMAC} function similar to traditional admission control mechanisms, while a smaller value leads to accepting more sessions and compromising QoE. We propose a model for $\beta$ in Section \ref{sec:beta}. 

A new requested session will be accepted by \emph{Pro-IBMAC} if the condition in Equation (\ref{eq:8}) meets:

\begin{equation}
\label{eq:8}
Pro\textit{-}IAAR(t) + x_{new} \le  C_{l} %
\end{equation}

Substituting Equations (\ref{eq:5}) and (\ref{eq:7}) in Equation (\ref{eq:6}), then Equation (\ref{eq:6}) in Equation (\ref{eq:8}), we get:

\begin{equation}
\label{eq:9}
\sum_{i=1}^n x_{i}(t)\  p_{i}(t)  \{ 1 + \beta (n-1) \}   + x_{new} \le  C_{l} %
\end{equation}

In Equation (\ref{eq:9}), \emph{x}$_{new}$ is the required rate of new session and \emph{C}$_{\emph{l}}$ is the link capacity. Studies recommend that peak rate be measured for \emph{x}$_{new}$ using techniques such as token buckets and traffic envelopes \cite{Floyd1996} and \cite{Qiu2001}. Others compute the peak rate of a new incoming flow by tracking the first \emph{A} packets of the flow and using sliding window \cite{Ammar2012}.
 
In summary, \emph{Pro-IBMAC} in Equation (\ref{eq:9}) employs \emph{Pro-IAAR(t)} in Equation (\ref{eq:6}) which is based on the Hoeffding inequality theorem. The value of $\gamma$ in Equation (\ref{eq:4}) specifies the level of optimization achieved by considering \emph{Pro-IAAR(t)} in terms of number of sessions that can be fitted on a particular link compared to the \emph{CalR($\tau$)} in Equation (\ref{eq:2}). 

\subsection{Proposed Model for $\beta$}
\label{sec:beta}

The tuning parameter $\beta$ affects the operation of the proposed algorithm. The value can be set to optimize the trade-off relationship between QoE of enrolled sessions and number of sessions. In this section, we develop a model for $\beta$. We estimate the value of $\beta$ using two publicly available video sequences; a 30-seconds clip called Mother And Daughter (\emph{MAD}) and a 35-seconds clip called \emph{Paris}. These two video sequences are used to validate the proposed $\beta$ model for various video content. Similar short sequences have also been used for video streaming service and subjective tests \cite{Khan2012}.

While choosing the videos, the following points were taken into consideration: firstly, long video is not practical for subjective tests in which subjects evaluate a numbers of videos. Secondly, the aim was to evaluate the admission control-specifically the acceptance/rejection of sessions-and evaluate the admission rate. Thus the duration of video is not expected to have effect on the evaluation of the proposed algorithm. The \emph{MAD} sequence was taken as a slow moving content due to low motion of its video scenes and \emph{Paris} as a fast moving content due to fast motion of its video scenes. The considered contents were classified into slow and fast based on common conventions and the size of their encoded frames, as faster content has larger frame size. Other studies have classified video contents in a similar way, e.g. \cite{Khan2012}. Details about the video sequences are shown in Table \ref{tab:video}. Other simulation settings including the coding and network parameters are explained in Section \ref{sec:simulationSetup}.

\begin{table}[!th]
\centering
\caption{Video sequence description}
\label{tab:video}
\begin{tabular}{ |l|l|l| }
\hline
 Description & Video sequence 1 & Video sequence 2 \\ \hline
 Name & \emph{MAD} & \emph{Paris} \\
 \hline
 Description & A mother and daughter & A woman playing with a \\
             & speaking, low motion. & ball and a man spinning a \\
             & & pen continuously, high \\
             & & motion. \\
             \hline
 Frame Size & CIF (352x288) & CIF (352x288) \\
 \hline
 Duration(second) & 30 & 35  \\ 
 \hline
 Number of frames & 900 & 1065 \\
  \hline
\end{tabular}
\end{table}

\begin{figure}[!th]
\centering
\includegraphics[width=\columnwidth, height=6.5cm]{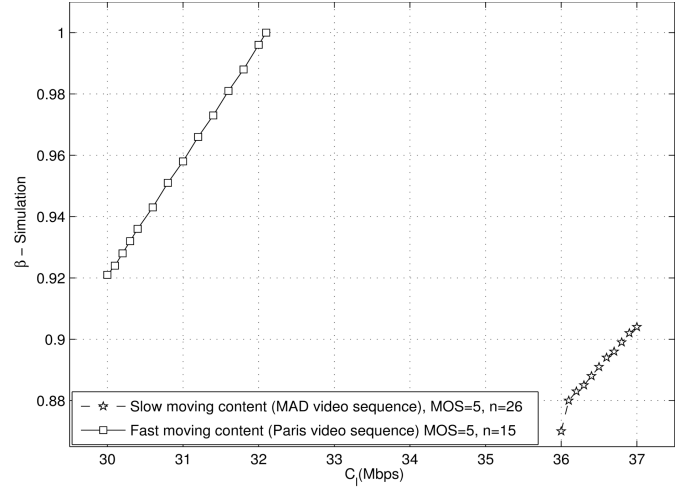}
\caption{$\beta$ - Link capacity relationship}
\label{beta_bw}
\end{figure}

\begin{figure}[!th]
\centering
\includegraphics[width=\columnwidth, height=6.5cm]{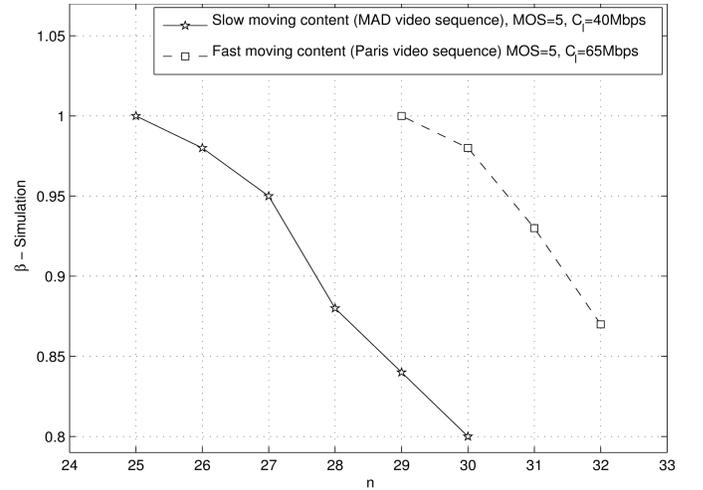}
\caption{$\beta$ - Number of sessions relationship}
\label{beta_sessions}
\end{figure}

\begin{figure}[!th]
\centering
\includegraphics[width=\columnwidth, height=6.5cm]{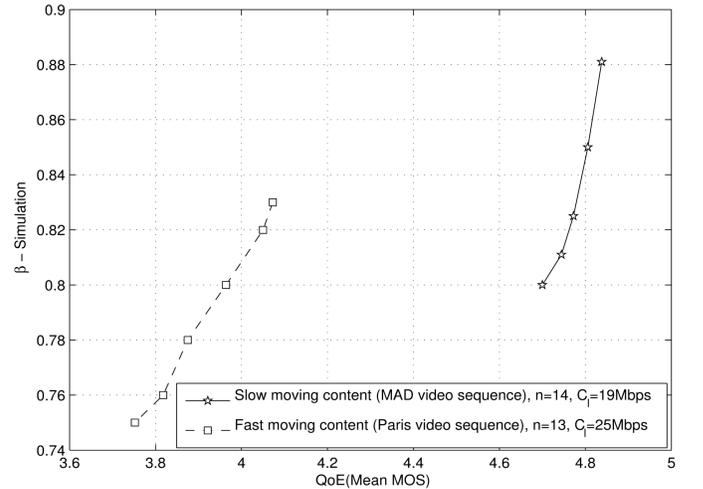}
\caption{$\beta$ - \emph{QoE} relationship}
\label{beta_mos}
\end{figure}

We run extensive simulation to find parameters that potentially affect $\beta$. \emph{$C_{l}$}, \emph{n} and \emph{QoE} were found to have impact on $\beta$. \emph{QoE} was measured by simulation which will be explained in Section \ref{sec:simulationSetup}. To understand the impact of any of these parameters on $\beta$, the values of the other two parameters (controlling parameter) were kept fixed. The values of the controlling parameters for both sequences are shown in Fig. \ref{beta_bw}, \ref{beta_sessions} and \ref{beta_mos}. These figures also show the relationship between $\beta$ and each of \emph{$C_{l}$}, \emph{n} and \emph{QoE} respectively. 

Equation (\ref{eq:10}) shows the mathematical relationship between the four parameters. However, in this paper we focus on a value of $\beta$ that produces excellent quality (MOS=5) only. Thus \emph{QoE} was not considered as a variable in the proposed model of $\beta$. The exponential relationship between $\beta$ and QoE shown in Fig. \ref{beta_mos} will be included to the model of $\beta$ in future studies to provide multi-class MOS.

\begin{equation}
\label{eq:10}
\beta \propto \frac{QoE,C_{l}}{n}
\end{equation}

The simulation data was analyzed with 2-way repeated analysis of variance (ANOVA) \cite{Miller1997} to confirm the significance of \emph{$C_{l}$} and \emph{n} in modeling of $\beta$. Also, it can find the difference between means given by the remaining two parameters \emph{$C_{l}$} and \emph{n}. ANOVA let us understand the effect of parameters and their interaction on $\beta$ which will later be used in the regression modeling. The ANOVA results are shown in Table \ref{table:anova} for F and p-values: the Cumulative Distribution Function (CDF) of F. Parameter with (p$<$0.05) is considered to have significant impact on $\beta$. The analysis results indicate that $\beta$ is affected by each of \emph{$C_{l}$} and \emph{n} as p-values are 0 and 0.0023 respectively. The result also shows that the combined parameters have no interaction effect on $\beta$ because the p-value is 0.6249. This can be justified by the fact that \emph{n} is determined by \emph{$C_{l}$}; the higher capacity of the link, the more sessions are accepted. Based on the value of p in the table, we can conclude that $\beta$ is affected more by \emph{$C_{l}$} than by \emph{n}.

The relationship between $\beta$, \emph{n} and \emph{C}$_{l}$ can be established from ANOVA analysis and Fig. \ref{beta_bw} and \ref{beta_sessions}. We found that there is a linear relationship between $\beta$ and \emph{C}$_{l}$ and a polynomial relationship between $\beta$ and \emph{n}. Finally, the rational model shown in Equation (\ref{eq:11}) was formulated to estimate the value of $\beta$ from the nonlinear regression analysis of the simulation data using MATLAB. The values of the coefficients of Equation (\ref{eq:11}) are listed in Table \ref{table:beta_validation_coefficient_slowMovingContents} and \ref{table:beta_validation_coefficient_fastMovingContents}. As \emph{n} is determined by the size of video frames (content dependent), different values for the model coefficients were found for slow (\emph{MAD} sequence) and fast (\emph{Paris} sequence) moving contents. The table also shows the correlation coefficient (R$^2$) and Root Mean Squared Error (RMSE) of the proposed model for both contents.

\begin{table}[!th]
\centering
\caption{ANOVA Results for Main and Interaction Effects}
\label{table:anova}
\begin{tabular}{|c|c|c|c|c|c|}
                                                                         \hline
Source & Sum of  & Degree of & Mean    & F & p-Value	\\
       & squares & freedom   & Squares &   & \\
\hline
C$_{l}$ 			& 0.33001 	& 1 & 0.33001 & 720.02 & 0 \\\hline
n	& 0.01807 	& 2 & 0.00903 & 19.71 & 0.0023 \\ \hline
C$_{l}$*n	& 0.00047 	& 2 & 0.00023 & 0.51 & 0.6249 \\ \hline
\end{tabular}
\end{table}

\begin{equation}   
\label{eq:11}\
\beta = \alpha + ( \frac{ C_{l} }{\delta * n} )
\end{equation}

\begin{table}[!th]
\centering
\caption{Coefficients of $\beta$ prediction model and model validation correlation coefficients - slow moving content (\emph{MAD} video sequence)}
\label{table:beta_validation_coefficient_slowMovingContents}
\begin{tabular}{c|c|c}
                                                                        \hline
 \multicolumn{2}{|c|}{$\alpha$} & \multicolumn{1}{c|}{$\delta$}  \\
\hline
\multicolumn{2}{|c|}{-0.5429} & \multicolumn{1}{c|}{0.9689}  \\\hline
\multicolumn{2}{c}{} &  \\\hline
\multicolumn{2}{|l|}{Adjusted R$^2$ (Validation)}  & \multicolumn{1}{c|}{\%88.44} \\\hline
\multicolumn{2}{|l|}{RMSE (Validation)} & \multicolumn{1}{c|}{0.0149} \\\hline
\end{tabular}
\end{table}

\begin{table}[!th]
\centering
\caption{Coefficients of $\beta$ prediction model and model validation correlation coefficients - fast moving content (\emph{Paris} video sequence)}
\label{table:beta_validation_coefficient_fastMovingContents}
\begin{tabular}{c|c|c}
                                                                        \hline
 \multicolumn{2}{|c|}{$\alpha$} & \multicolumn{1}{c|}{$\delta$}  \\
\hline
\multicolumn{2}{|c|}{-0.1227} & \multicolumn{1}{c|}{1.952}  \\\hline
\multicolumn{2}{c}{} &  \\\hline
\multicolumn{2}{|l|}{Adjusted R$^2$ (Validation)}  & \multicolumn{1}{c|}{\%90.54} \\\hline
\multicolumn{2}{|l|}{RMSE (Validation)} & \multicolumn{1}{c|}{0.0124} \\\hline
\end{tabular}
\end{table}

The model for $\beta$ was proposed based on two video sequences (\emph{MAD} and \emph{Paris}), however the methodology is similar and applies to faster moving content, such as sports, of the same format (CIF). Thus, the model is limited to the video format and coding parameters specified in Table \ref{tab:video}. The model can be applied to other formats and coding parameters with different coefficient values. This is because other formats and/or coding parameters generate different frame sizes and bit rates which control the number of sessions (parameter \emph{n} in the model) for a specific link capacity (parameter C$_{l}$ in the model). They only have impact on the value of the coefficients of the model. The model will be validated by CIF and QCIF video formats in Section \ref{sec:validation}.

\section{Results and Analysis}
\label{sec:resultANDdiscuttion}

The simulation environment is explained in Section \ref{sec:simulationSetup}. Section \ref{sec:comparison} compares the proposed \emph{Pro-IBMAC} to \emph{CBAC} in terms of MOS and number of the sessions, packet drop ratio and delay. The impact of $\beta$ on the functionality of \emph{Pro-IBMAC} is discussed in Section \ref{sec:beta_Impact}. 

\subsection{Simulation Setup}
\label{sec:simulationSetup}

Since the number of admitted sessions for a specific link capacity is the target of this study, only acceptance/rejection policy of admission control was investigated. Queue size and simulation time were chosen so as not to cause drop due to insufficient queue length or time. The video format such as CIF or QCIF has impact on the number of admitted sessions due to the difference in the size of encoded frames. In this paper, CIF (352x288) is assumed for input video as an acceptable video format for most video capable devices such as handsets and mobiles \cite{Khan2012}. It is also suitable for videoconferencing systems delivered on telephone lines. While modern devices support much higher resolution, CIF makes packet level simulation practical. A bottleneck link of dumbbell topology over which the video sources send to their peer destinations was considered for the implementation of the proposed \emph{Pro-IBMAC} scheme. In addition to $\beta$, \emph{$C_{l}$} was the main variable in the simulation. Other parameters such as link delay, queue length, packet size were kept fixed. Lost packets were replaced with 0 by the etmp4 \cite{Gross2004} decoder as a way for coping with the losses. The values of the simulation parameters and settings are shown in Table \ref{tab:parameter_settings}.

\begin{table}[!th]
\centering
\caption{Encoder and Network Settings}
\label{tab:parameter_settings}
\begin{tabular}{ |l|l|l| }
\hline
 & Parameter & Value \\ \hline
\multirow{3}{*}{Encoder} & Frame size & CIF(352x288) \\
 & Frame rate & 30fps \\
 & Group of picture & 30 \\ \hline
\multirow{8}{*}{Network} & \emph{$C_{l}$}(Mbps) & 22, 24, 30, 36, 39, 40 \\
 & Topology & Dumbbell \\
 & Packet size(byte) & 1024 \\
 & UDP header size(byte) & 8 \\
 & IP Header Size(byte) & 20 \\
 & Queue size(packet) & 5300 \\
 & Queue management algorithm & Droptail \\ 
 & Queue discipline & FIFO(First In First Out) \\ 
 & Simulation time(second) & 500 \\ 
\hline
\end{tabular}
\end{table}

New sessions were requested randomly and continuously every second. They were accepted as long as there was enough bandwidth on the bottleneck link i.e: Equation (\ref{eq:9}) was satisfied. NS-2 \cite{ns2} was used to measure \emph{CalR($\tau$)} and \emph{Pro-IAAR} and implement \emph{CBAC} and \emph{Pro-IBMAC}. The implementation of the proposed \emph{Pro-IBMAC} is summarized in Algorithm \ref{pro-ibmac}.

\begin{algorithm}[h]
\caption{Proposed \emph{Pro-IBMAC}}
\label{pro-ibmac}  	
\begin{algorithmic}

\Statex $Given\ \emph{$C_{l}$},\ x_{new}\ and\ \emph{n}$

\For  {$Every\ video\ session\ request$} 
\Statex $Compute\ \mu_S(t)\ from\ Equation\ (\ref{eq:5})$
\Statex $Compute\ \beta\ from\ Equation\ (\ref{eq:11})$	
\Statex $Compute\ Pro\textit{-}IAAR(t)\ from\ Equation\ (\ref{eq:6})$

\If{  $Equation\ (\ref{eq:9}) = True$ } 
\State $Request\ accepted$ 
\Else
\State $Request\ rejected$ 
\EndIf  
\EndFor

\end{algorithmic}
\end{algorithm}

The time window $\tau$ has an impact on the operation of the admission control. The smaller the value of $\tau$, the more conservative the admission control and more sensitive to the traffic bursts. On the other hand, the larger the value of $\tau$, the smoother the measured rate and less reactive to the changes in the network load. In practice, $\tau$ will be a few seconds \cite{Latre2011}. In this paper, \emph{IAAR(t)} was averaged over 1-second.

The \emph{MAD} video sequence described in Table \ref{tab:video} was fed to the NS-2 simulator using EvalVid \cite{Gross2004}. Evalvid provides a set of tools to analyze and evaluate video quality by means of PSNR and MOS metrics. The Evalvid MOS metric (We call it simulated MOS) was used in this paper which calculates the average MOS value of all frames for the entire video with a number between 1 and 5, instead of the frame-wise PSNR metric. The MOS metric represents the impression of the user for the entire received video and has been widely used by research community \cite{Li2010,Khan2010a,Kim2012,Khan2009a,Tommasi2014,Papadimitriou2007,Khan2009,Ma2012,Aguiar2008,Erdelj2013,Tan2013,Escuer2014}. Although the MOS metric does not map very well to the subjective impression for a long video sequence, it was used for short video sequences (30-35 seconds) in this paper. In addition to the MOS metric, we calculated the Distortion In Interval (DIV) metric \cite{Gross2004} to restrict the MOS metric within a fixed interval (30 frames in this paper). This stringent metric calculates the maximum percentage of received frames with a MOS smaller than that of the sent frame within a given interval. 

The efficiency of the proposed \emph{Pro-IBMAC} and \emph{CBAC} was evaluated based on MOS, number of sessions, packet drop ratio and mean delay. These performance metrics were chosen due to their impact on multimedia traffic. The performance of \emph{Pro-IBMAC} was tested to find the maximum number of video sessions on a bottleneck link while keeping the QoE of each session at acceptable or required levels. This was compared to other procedures such as \emph{CBAC}. The objective was to see how \emph{Pro-IBMAC} utilizes the available bandwidth compared to \emph{CBAC}. Further simulations were used to investigate the effect of parameter $\beta$ on the performance metrics.

\subsection{\emph{Pro-IBMAC} vs \emph{CBAC}}
\label{sec:comparison}
  
\begin{figure}[!th]
\centering
+-\includegraphics[width=\columnwidth, height=5cm]{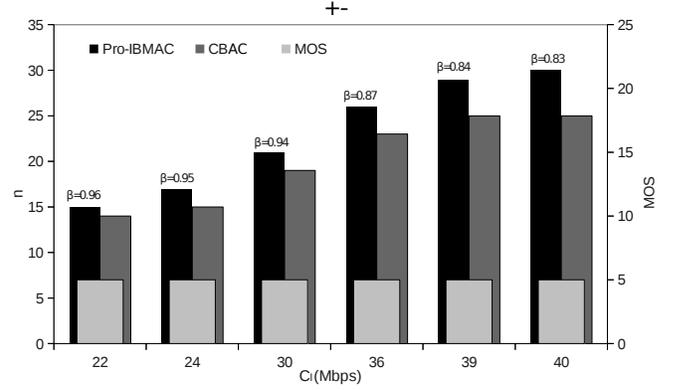}
\caption{MOS of the \emph{CBAC} and \emph{Pro-IBMAC} admitted sessions}
\label{session_mos}
\end{figure}

It has been found that there is a considerable difference between the two schemes in terms of the number of accepted sessions. This is plotted in Fig. \ref{session_mos}. The number of admitted sessions is always higher for \emph{Pro-IBMAC}. The difference between the number of admitted sessions increases with increasing of the link capacity. For example, the number of admitted sessions to 22Mbps link is 15 against 14 for \emph{Pro-IBMAC} and \emph{CBAC} respectively, whereas it is 30 against 25 in the case of 40Mbps link. The main role of any admission control is to ensure that the acceptance of a new session does not violate the QoE of on-going sessions. We computed the MOS of every single accepted session for both schemes. We found that increase in \emph{n} does not come at the cost of QoE as all accepted sessions by \emph{Pro-IBMAC} and \emph{CBAC} were scored MOS 5. Note that the MOS of video sessions is labeled on the secondary y-axis in the figure. The value of $\beta$ that produces this increase in \emph{n} and guarantees the video quality is also shown in the figure. This will be further described in Section \ref{sec:beta_Impact}. However this simulation outcome can not be generalized. \emph{Pro-IBMAC} may not guarantee the same level of QoE as \emph{CBAC} in a real implementation. This is because our proposed scheme is based on a probabilistic approach therefore, there is a possibility of the upper bound to be lower than the bursty instantaneous rate, especially for small $\tau$.

Table \ref{tab:packetdropratio_BW} shows mean MOS and DIV. The DIV values (0\%) indicate that all received frames have the same MOS as of the original frames. It also lists the packet drop ratio of the accepted sessions of \emph{Pro-IBMAC} and \emph{CBAC} for each link. Since we aim at a $\beta$ value that doesn't degrade the MOS of received videos, as mentioned in Section \ref{sec:beta}, no packet drop was expected.

\begin{table}[!th]
\centering
\caption{Packet drop ratio and admitted sessions of \emph{Pro-IBMAC} and \emph{CBAC}}
\label{tab:packetdropratio_BW}
\begin{tabular}{|c|c|c|c|c|c|c|}
                                                                         \hline
\multicolumn{4}{|c|}{{}} & \multicolumn{2}{c|}{{\emph{Pro-IBMAC}}}  &    	\emph{CBAC} \\\hline    
\emph{$C_{l}$}(Mbps)  & Packet & & & &  & \\
  & Drop & MOS & DIV & $\beta$ & \emph{n} & \emph{n}\\
  & \% & & \% & &  & \\
\hline
22 & 0 & 5 & 0 & 0.96 & 15 &  14 \\\hline
24 & 0 & 5 & 0 & 0.95 &  17 &  15 \\\hline
30 & 0 & 5 & 0 & 0.94 &  21 &  19 \\\hline
36 & 0 & 5 & 0 & 0.87 &  26 &  23 \\\hline
39 & 0 & 5 & 0 & 0.84 &  29 &  25 \\\hline
40 & 0 & 5 & 0 & 0.83 &  30 &  25 \\\hline
\end{tabular}
\end{table}

\begin{figure}[!th]
\centering
\includegraphics[width=\columnwidth]{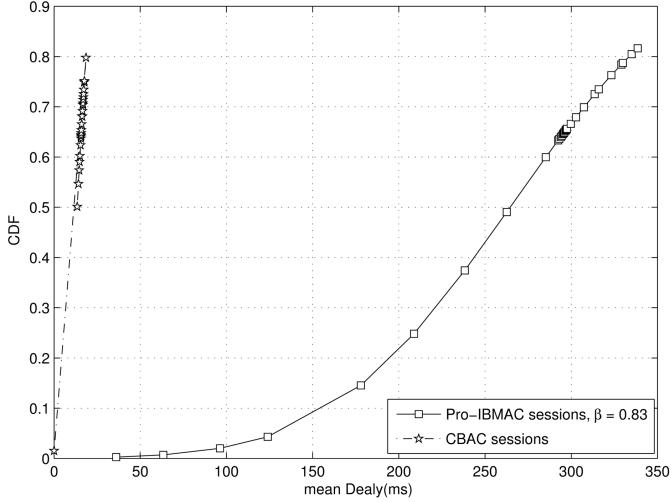}
\caption{CDF of the mean delay of the \emph{CBAC} and \emph{Pro-IBMAC} sessions}
\label{meanDelay_both_CDF}
\end{figure}

As for the delay, we measured the mean delay using the ns-2 trace files for both schemes. Fig. \ref{meanDelay_both_CDF} illustrates the CDF of the mean delay for the \emph{Pro-IBMAC} and \emph{CBAC} sessions for 40Mbps link. As shown in Table \ref{tab:packetdropratio_BW}, 30 sessions are accepted by \emph{Pro-IBMAC} for $\beta$=0.83 and 25 by \emph{CBAC}. More sessions on the same link by \emph{Pro-IBMAC} caused higher delay due to more buffering. Therefore the \emph{Pro-IBMAC} sessions experienced higher delay compared to the lower delay of the \emph{CBAC} sessions. Nevertheless, increase in the delay that comes at the cost of the optimization of QoE-Session can not be tolerated by real-time video traffic. For \emph{Pro-IBMAC} to be applicable to realtime traffic, a proper value of $\beta$ must be selected. Video streaming services can tolerate a delay of 5-seconds \cite{Szigeti2004}, thus it can be used within this limit. In future work, we will further investigate this relationship and develop the model of $\beta$ to include delay as another variable.

\subsection{The Impact of $\beta$ on \emph{Pro-IBMAC}}
\label{sec:beta_Impact}
\begin{figure}[!th]
\centering
\includegraphics[width=\columnwidth]{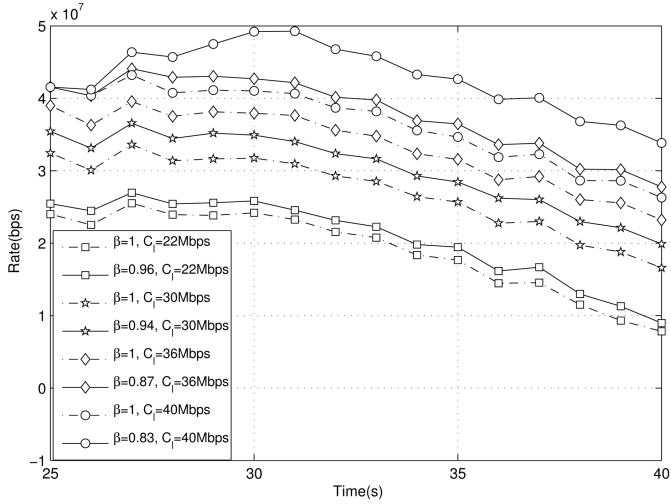}
\caption{IAAR and upper limit of the exceedable rate for different link capacities}
\label{rate_all}
\end{figure}

\begin{figure}[!th]
\centering
\includegraphics[width=\columnwidth]{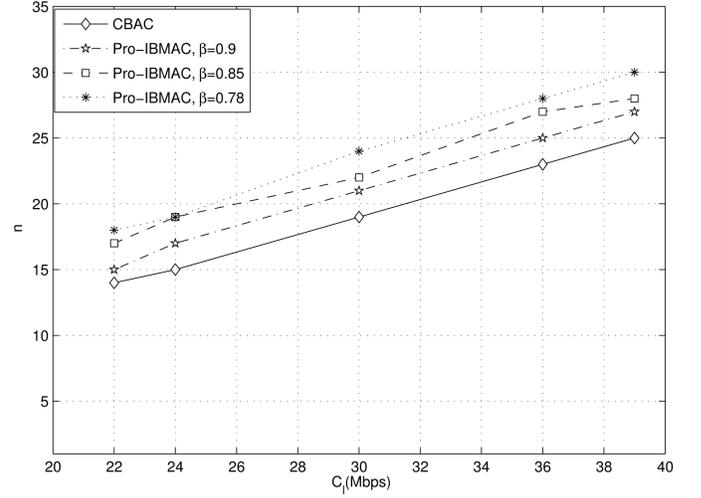}
\caption{Admitted sessions of \emph{CBAC} and \emph{Pro-IBMAC} for different link capacities}
\label{session_bandwidth}
\end{figure}

As mentioned earlier, parameter $\beta$ controls the degree of risk between the admission decision and QoE of existing sessions. Fig. \ref{rate_all} shows \emph{IAAR(t)} (dash-dot line) and the upper limit of the exceedable aggregate rate (solid line) that allows more sessions (compared to sessions allowed by \emph{IAAR(t)}), without QoE degradation of enrolled video sessions. The proposed value of $\beta$ for four scenarios (22, 30, 36 and 40Mbps) is shown in the figure. It can be seen that the lower the value of $\beta$, the wider the gap between the two rates. Decreasing $\beta$ causes increase in the limit of the exceedable rate. This makes \emph{Pro-IBMAC} more flexible and it accepts more sessions. This can be better observed in Fig. \ref{session_bandwidth}. It depicts the number of admitted sessions for different link scenarios. The solid line shows the number of sessions admitted by \emph{CBAC}, while the other three lines show sessions admitted by \emph{Pro-IBMAC} for three different value of $\beta$ (0.9, 0.85 and 0.78). For the same link, the linear relationship between \emph{n} and \emph{$C_{l}$} allows more sessions to be accepted by lowering the value of $\beta$. For instance, for 39Mbps link, \emph{Pro-IBMAC} accommodates 27, 28 and 30 sessions for $\beta$=0.9, 0.85 and 0.78 respectively compared to 25 sessions of CBAC. Note that $\beta$ $\ge$ 0.84 guarantees accepted sessions with MOS 5 as shown in Table \ref{tab:packetdropratio_BW}.

\begin{figure}[!th]
\centering
\includegraphics[width=\columnwidth]{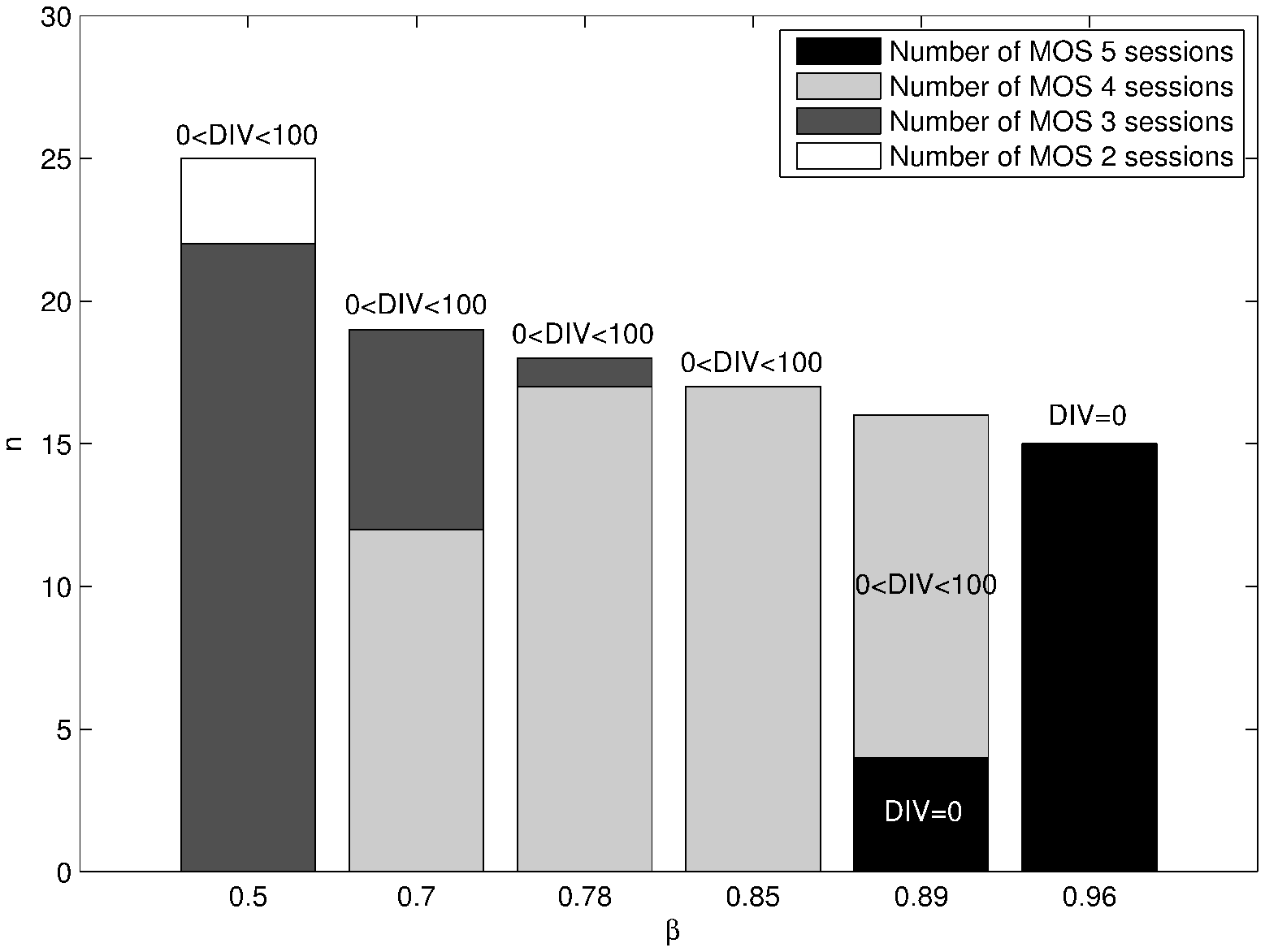}
\caption{Impact of $\beta$ on MOS and \emph{n}, \emph{$C_{l}$}=22Mbps}
\label{beta_MOS_sessions_22Mbps}
\end{figure}

However, continuous decreasing of $\beta$ will degrade the QoE of admitted sessions as more sessions are accepted. Therefore, care is required to fine tune the value of $\beta$ that optimizes the operation of \emph{Pro-IBMAC}. The aim is to accept as many sessions as possible, while keeping the QoE of the sessions at required levels. As per the proposed model, the value of $\beta$ depends on \emph{$C_{l}$}, \emph{n} and required \emph{QoE}. We investigated this further for 22Mbps and 24Mbps links. Fig. \ref{beta_MOS_sessions_22Mbps} shows the number of MOS 2, 3, 4 and 5 sessions separately as well as total number of sessions for 22Mbps link. If we consider that the required class of QoE is MOS 5, then the proposed value of $\beta$ is 0.96, i.e. for $\beta$ less than 0.96, sessions with multi-MOS levels exist, while for $\beta$ $\ge$ 0.96 all sessions score a MOS of 5. It also can be seen in Fig. \ref{beta_MOS_sessions_22Mbps} that decreasing $\beta$ from 0.96 to 0.5 increases the total number of sessions and number of MOS 3 and 2 sessions while decreasing the number of MOS 5 and 4 sessions.

\begin{figure}[!th]
\centering
\includegraphics[width=\columnwidth]{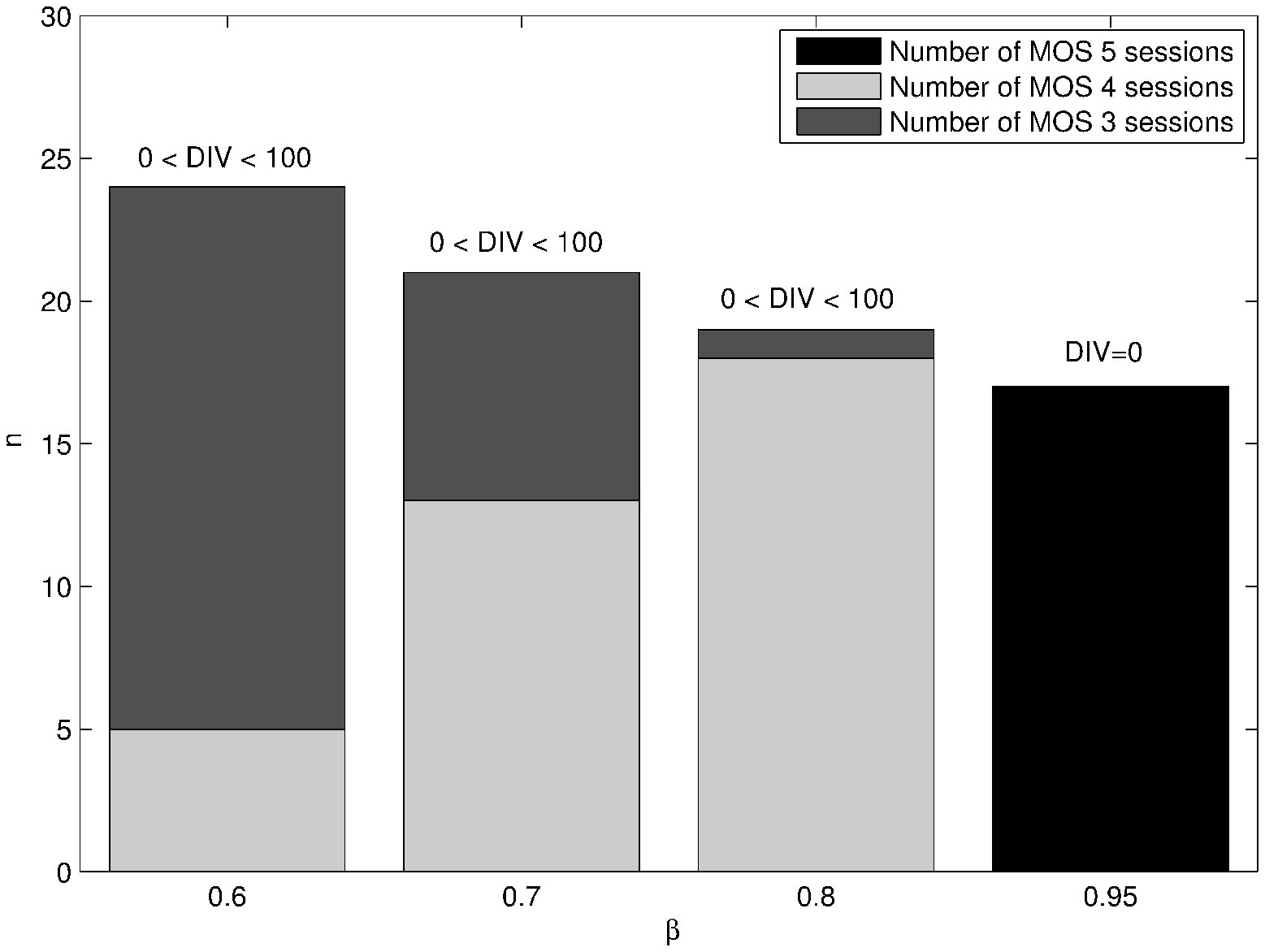}
\caption{Impact of $\beta$ on MOS and \emph{n}, \emph{$C_{l}$}=24Mbps}
\label{beta_MOS_sessions_24Mbps}
\end{figure}

In another scenario, we found that the proposed value of $\beta$ is 0.95 for 24Mbps link as shown in Fig. \ref{beta_MOS_sessions_24Mbps}. $\beta$ of 0.95 or greater, maintains the MOS of accepted sessions at 5, while $\beta$ less than 0.95 produces sessions with multi-MOS classes which comes at the cost of the QoE of enrolled sessions. For instance, $\beta$ of 0.8 creates 18 sessions with MOS 4 and 1 session with MOS 3. Whilst $\beta$ of 0.6 changes the number of MOS 4 sessions to 5 and MOS 3 sessions to 19 . Note that there are 19 sessions in total for $\beta$=0.8 and 24 sessions for $\beta$=0.6. 

Fig. \ref{beta_MOS_sessions_22Mbps} and \ref{beta_MOS_sessions_24Mbps} also show the DIV values of accepted sessions at different $\beta$ values. As the DIV was 0\% for sessions with MOS 5 and between 0\% and 100\% for sessions with MOS\textless5, in the figures we simply labeled DIV=0 to denote all the accepted sessions are MOS 5 and 0\textless DIV\textless100 denote that MOS of sessions are less than 5.

Although most real-time applications can tolerate some packet loss, more than an acceptable level may degrade the quality of received video. As expected, fewer sessions of \emph{CBAC} will guarantee no packet loss, in contrast extra added sessions of \emph{Pro-IBMAC} cause packet drop when $\beta$ is set lower than the proposed value and increases slightly with the increase of the number of sessions. Table \ref{table:packetdropratio_C} presents the percentage of the packet drop ratio of the \emph{Pro-IBMAC} admitted sessions for different value of $\beta$ for 22Mbps link. The ratio increases with the decrease of $\beta$ due to fitting a higher number of sessions into the same link. The table shows 0.45\%, 4.06\% and 6.70\% drop of the total number of packets for $\beta$= 0.89, 0.85 and 0.78 respectively. The proposed value of $\beta$ (0.96) ensures that no packets are dropped as shown in the table. 

\begin{table}[!th]
\centering
\caption{Packet drop ratio and admitted session of \emph{Pro-IBMAC} for different $\beta$, \emph{C}$_{\emph{l}}$ = 22Mbps}
\label{table:packetdropratio_C}
\begin{tabular}{|c|c|c|}
                                                                         \hline
$\beta$ & Packet Drop Ratio \% & \emph{n}\\
\hline
0.96 & 0 & 15 \\\hline
0.89 & 0.45 & 16 \\\hline
0.85 & 4.06 & 17 \\\hline
0.78 & 6.70 & 18 \\\hline
\end{tabular}
\end{table}

Improper values of $\beta$ not only causes packet drop, but it also degrades the MOS levels (discussed earlier) and increases the delay. Fig. \ref{meanDelay_both_CDF} demonstrates how a high number of sessions caused by a low value of $\beta$ can contribute to the increase of the delay which can be substantial for a large number of sessions.

The disadvantage of lowering the value of $\beta$ is not only that it causes degradation to the MOS level of video sessions, or increase in the delay and packet loss. We observed that the decoder takes longer to decode and play back the received video for low value of $\beta$, for instance when $\beta$=0.6 for 24Mbps link. The ISP can tune the value of $\beta$ to control the trade-off between providing the required level of QoE and increasing their revenue by accommodating as many user sessions as possible.

\section{Subjective Tests}
\label{sec:subjectiveTest}

We performed subjective tests to involve human subjects in rating the quality of videos. The tests followed the ITU-R BT.500-13 recommendation \cite{ITU-R2012}. The five-grade scale from 1 to 5 of the Single Stimulus (SS) adjectival categorical judgment method was used in which 1 represents 'bad' and 5 represents 'excellent' quality. Each video was presented in random order and rated individually by 17 subjects one at a time. The number of participants exceeded the minimum recommended number (15 subjects).

As the \emph{MAD} sequence was chosen, 48 videos delivered through different link capacities and different values of $\beta$ shown in Table \ref{tab:packetdropratio_BW}, Fig. \ref{beta_MOS_sessions_22Mbps} and Fig. \ref{beta_MOS_sessions_24Mbps} were used in the tests. They were decoded from the simulations and selected from Fig. \ref{session_mos} (MOS 5), Fig. \ref{beta_MOS_sessions_22Mbps} (MOS 2, 3, 4 and 5) and Fig. \ref{beta_MOS_sessions_24Mbps} (MOS 3, 4 and 5). The description of the testing video sequence, coding and network parameters were the same as described in Tables \ref{tab:video} and \ref{tab:parameter_settings}. Each video was identified by the MOS value calculated with Evalvid, regardless of the capacity of the link and/or value of $\beta$. The aim was to have variety of videos with different MOS values through changing the capacity of the link and value of $\beta$. The simulated $\beta$ and predicted $\beta$ of the testing videos will be plotted in Section \ref{sec:validation}.

The videos were presented in their original size (352x288), embedded in a separate web page with grey background and rated on the same page. There were two sessions, each lasting up to 30 minutes with 10 minutes break in between. To stabilize the subjects' opinion, five dummy videos were displayed at the beginning of the session without considering their scores. Prior to the actual rating, the subjects were carefully introduced to the assessment method, likely quality artifacts that might be observed, rating scale and timing. They were given unrestricted time and the viewing distance was comfortable. 

The tests were conducted in a white background laboratory on 29 inch LCD monitor (Dell P2213) with 1680x1050 resolution and 32 bit true color. 5 female and 12 male non-expert observers participated in the tests. All participants were university students, 1 in the range of 18-25, 7 in the range of 26-30 and 9 over 30. At the end of the tests, subjects who were surveyed on the duration and comfortability of the tests did not express any concern. The subjects were screened for any possible outliers following the screening procedure of the SS method \cite{ITU-R2012}. Two subjects have been eliminated and their data were not considered in the analysis. The MOS was calculated by taking the mean score for each of the videos following the procedure described in \cite{ITU-R2012}.

\begin{figure}[!th]
\centering
\includegraphics[width=\columnwidth]{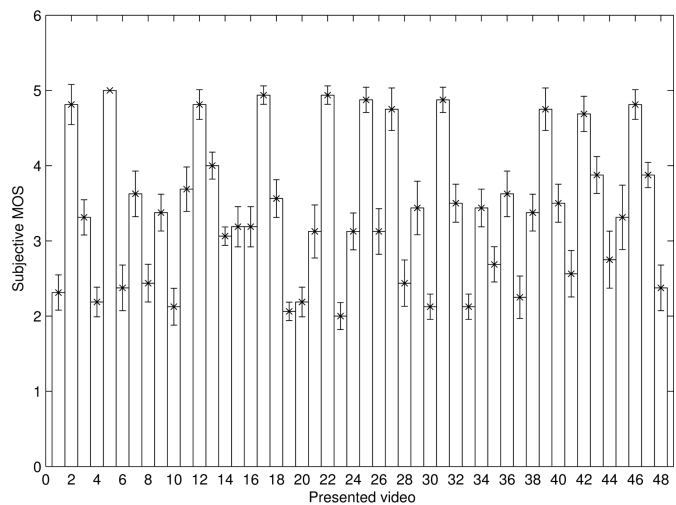}
\caption{Bar chart of the subjective MOS with confidence interval for individual video}
\label{fig:scores_percentage_video}
\end{figure}

\begin{figure}[!th]
\centering
\includegraphics[width=\columnwidth]{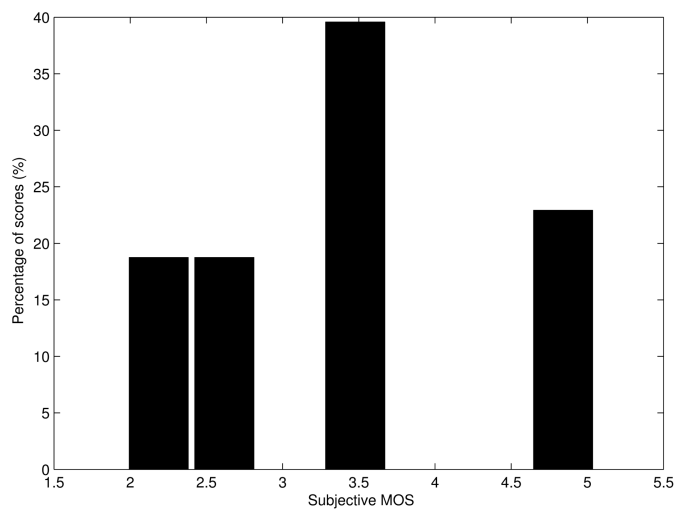}
\caption{Bar chart of the percentage of scores of the subjective MOS}
\label{fig:scores_percentage_mos}
\end{figure}

The bar chart in Fig. \ref{fig:scores_percentage_video} illustrates the subjective mean MOS of every presented video with the confidence interval. It shows the mean and range (the upper and lower limits) of MOS given to each video by the subjects. The analysis shows that around 40\% of the scores went for a MOS of 3.5. The distribution of the scores is plotted in Fig. \ref{fig:scores_percentage_mos}.

\section{Validation of the Proposed Model}
\label{sec:validation}

In this section, the validation of the proposed model of $\beta$ with simulation results is explained. It also demonstrates the validation of the simulated MOS with subjective MOS.

\begin{figure}[!th]
\centering
\includegraphics[width=\linewidth]{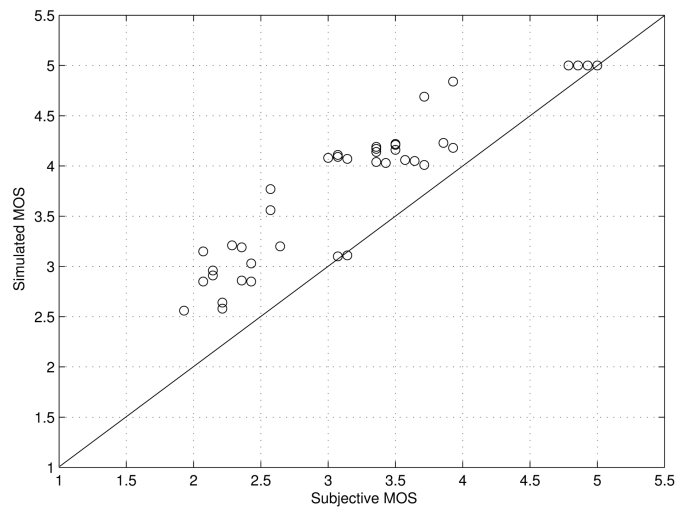}
\caption{Validation of the simulated MOS with subjective MOS}
\label{mos_validation}
\end{figure}

The scatter plot in Fig. \ref{mos_validation} shows the simulated MOS against subjective MOS. Overall, the subjects were irritated by video impairments, their scores therefore underestimate the simulation scores. The majority of simulated MOS scores seen are higher than subjective MOS. However, both scores are getting closer for less impaired videos (subjective MOS between 4.78-5). These videos were delivered with the proposed values of $\beta$ for each value of \emph{$C_{l}$}. Note that as there are about 11 overlapping scores within this range, all can not be seen in the figure. Overlapping of the scores can be further noticed in Fig. \ref{fig:scores_percentage_video}, in which there are 11 scores in the range of 4.78-5. The relationship is nearly linear correlated for videos delivered with the proposed value of $\beta$ that have MOS close to 5. This indicates that the model can provide better quality for end users with the proposed value of $\beta$.

\begin{figure}[!th]
\centering
\includegraphics[width=\columnwidth]{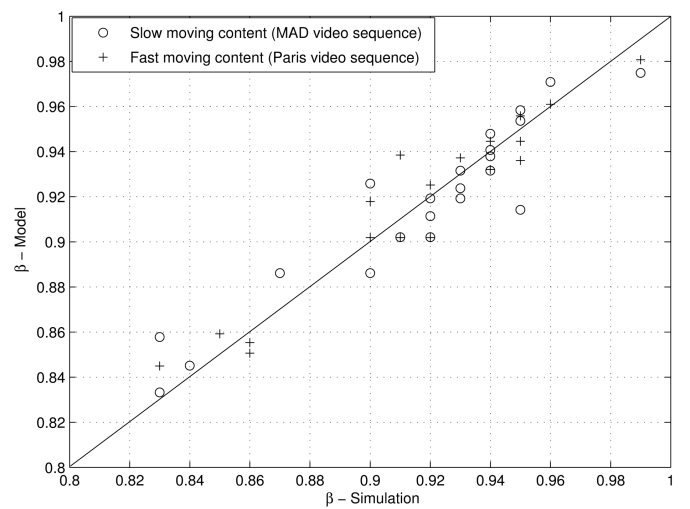}
\caption{Validation of the proposed model of $\beta$ with simulation results}
\label{beta_validation_without_MOS_Separate_cif}
\end{figure}

$\beta$ predicted by the model (Equation \ref{eq:11}) has been validated by the one found by simulations. Fig. \ref{beta_validation_without_MOS_Separate_cif} shows the resulting $\beta$'s scatter point plot of the predicted $\beta$ against simulated $\beta$ for slow and fast moving contents separately. As shown in Tables \ref{table:beta_validation_coefficient_slowMovingContents} and  \ref{table:beta_validation_coefficient_fastMovingContents}, the model of $\beta$ suits fast moving content with a correlation coefficient of 90.54\% compared to 88.44\% for slow moving content. This can be also observed in Fig. \ref{beta_validation_without_MOS_Separate_cif}. Thus, the model best suits dynamic content with high variation in bitrate. Note that there were few videos for each value of $\beta$ plotted in the figure, therefore the number of plotted points is less than the number of the testing videos (48).

As mentioned in Section \ref{sec:beta}, the model of $\beta$ can be applied to other video formats with different values of coefficients $\alpha$ and $\delta$. It has been validated by QCIF video format using the 45-seconds \emph{Deadline} video sequence of 1374 frames. The model achieved an adjusted R$^2$ of 83.59\% and RMSE of 0.0194. The values of $\alpha$ and $\delta$ were -0.1323 and 0.4991 respectively.

\section{Conclusion}
\label{sec:conclusion}
We proposed a novel algorithm to find the upper limit of the video total rate that can exceed a specific link capacity without QoE degradation of ongoing video sessions. A mathematical model for the measurement algorithm was developed and implemented in an admission control system. Its performance has been validated by simulating publicly available video sequences and subjective tests. The exceedable limit has been defined by parameter $\beta$ in the algorithm. This parameter can be used by ISPs to balance the trade-off between QoE and the number of video sessions. The simulation results have shown that the proposed admission control compared to calculated rate-based admission control optimizes the trade-off relationship between QoE-Session through fine tuning the value of $\beta$. The proposed algorithm can be applied within the scope of the video format and coding parameters specified in this paper. In future work, we will further develop the model of $\beta$ to include delay as another variable. The calculated MOS will be compared with SSIM metric. Moreover, an implementation of the proposed scheme in a cross-layer architecture for optimizing the QoE of video session will be investigated.

\section{Acknowledgment}
The authors would also like to thank volunteers for their participation in the subjective tests and the anonymous reviewers for their constructive comments.

\ifCLASSOPTIONcaptionsoff
  \newpage
\fi



\bibliographystyle{IEEEtran}

%

\end{document}